# Denouement of a wormhole-brane encounter

Enrico Rodrigo

*Department of General Studies, Charles Drew University*

*Los Angeles, CA 90059, USA*

Higher-dimensional black holes have long been considered within the context of brane worlds. Recently, it was shown that the brane-world ethos also permits the consideration of higher-dimensional wormholes. When such a wormhole, preexisting in the bulk, impinges upon our universe, taken to be a positive-tension 3-brane, it can induce the creation in our universe of a wormhole of ordinary dimensionality. The throat of this wormhole might fully constrict, pinch off, and thus birth a baby universe. Alternatively, the induced wormhole might persist. I show that persistence is more likely and note that the persistent wormhole manifests as a particle-like object whose interaction with cosmic matter is purely gravitational. I consider briefly the viability of this object as a dark matter candidate.

## I. INTRODUCTION

Wormholes have been discussed in the context of brane worlds ever since it was noticed [1] that the original Randall-Sundrum two-brane construction [2] meets the formal definition of a wormhole. In this construction the branes served as boundaries of a higher-dimensional spacetime – the bulk. This idea of branes as boundaries can be extended to bulk spacetimes that are not simply connected. In this case the bulk would contain holes that are bounded by closed branes. These holes would be regions of literal nothingness or void. The closed branes serving as boundaries between the bulk and this void may be modeled as the throats of thin-shell semi-wormholes [3].

A bulk containing such wormholes in addition to the brane defining our universe is not dissimilar to the oft-considered brane worlds featuring a bulk inhabited by one or more black holes (see [4] and references therein). Unlike the latter case, which implies the existence of singularities in the bulk, a brane world complimented by bulk-dwelling wormholes faithfully adheres to the canonical proscription of off-brane matter. For this reason the interaction of wormholes in the bulk with the brane defining our universe is at least as interesting a priori as the analogous interaction between our universe and bulk-dwelling black holes. The latter, which has been the subject of recent investigations [5, 6], is complicated by the existence of the event horizon of the black hole. Frolov's recent model of this interaction shows that the geometry of the brane is increasingly distorted by the approach of a higher-dimensional black hole, until it induces within the brane the formation of a new black hole, whose dimensionality is lower – matching that of the brane. This induced black hole forms, when the brane enters the horizon of the original bulk black hole.

Recently, it was suggested [7] that the wormhole-brane interaction is analogous to that between a brane and a black hole, with the role of the black hole's horizon being played by the wormhole's throat. The existence of a throat would seem to permit the envelopment of the incident wormhole by the brane on which it has impinged. The result of this envelopment, analogous to the aforementioned denouement of certain brane-black-hole encounters, would be the induced formation within the brane of a new wormhole. The purpose of this note is to consider the result of an encounter between a brane and a wormhole. I shall in particular address the question of whether *partial* envelopment of the wormhole by the brane – the condition corresponding to a persistent induced wormhole – necessarily proceeds to *total* envelopment. The latter is tantamount to the birth of a baby universe that occurs, when the throat of the induced wormhole becomes arbitrarily small.

If induced wormholes do not persists, but instead birth baby universes, we can rule them out as dark matter candidates. The motivated to consider such seemingly outré candidates, as those inferred from brane worlds, derives from the recent negative findings of the Cryogenic Dark Matter Search (CDMS) collaboration [23] to find evidence of conventional WIMPs.

## II. BRANE DESCRIPTION

In order to determine whether envelopment of a wormhole by a brane is in fact possible, we will consider the action of the static brane interacting gravitationally with a bulk wormhole. From the point of view of the bulk, the wormhole-brane system is not spherically symmetrical. Gravitational waves will therefore be emitted, as the wormhole impinges on the brane. This dissipative effect suggests that the configurations of the brane that minimize its static action are possible end states of a dynamic encounter. Such an encounter would in general require numerous bounces of the wormhole against the brane, before the final configuration predicted by the static action is reached. I shall make no attempt to model the dynamics of the encounter or to estimate the rate at which the local energy density is dissipated by gravitational waves. Nor shall I consider explosive or otherwise dissipative effects of brane-brane interaction through the emission within the macro brane (identified with our universe) of outbound fluxes of standard-model fields. Rather, I will focus on whether a conservatively defined static action permits total envelopment of the bulk wormhole by the brane. If it does not, we may then conclude that an induced wormhole persists -- that partial envelopment is not necessarily an intermediate step toward total envelopment and the formation of a baby universe.

We begin our detailed description of the static result of a wormhole-brane encounter by specifying the bulk wormhole. Let it be an *N*-dimensional, asymptotically anti-de Sitter Reissner-Nordstrøm black hole spacetime from which a hyper-cylindrical region, centered on *r=0* and enclosed by a "surface" with topology $S^{N-2}$ x $R^1$, has been excised.

That is to say, each spacelike slice (obtained by holding the time coordinate constant) will be missing a central region enclosed by an (*N-2*)-sphere. By choosing the radius of this sphere to exceed that of the event horizon of the corresponding (nonrotating) black hole, we ensure that the resulting geometry describes a (horizon-free) semi-wormhole. At the boundary (with topology $S^{N-2} \times R^1$) between this spacetime and the void created by the aforementioned excision – this boundary being an (*N-2*)-sphere in each *t=constant* spacelike slice -- we place the world tube of a closed negative-tension brane. This is in effect a spherically symmetrical thin-shell semi-wormhole, whose metric has the form

$$ds^2 = g_{\mu\nu} dX^\mu dX^\nu = -F(r) dt^2 + \frac{dr^2}{F(r)} + r^2 d\Omega_{N-2} \qquad (1)$$

with the restrictions $r \geq r_T$ and $r_T > r_H$, where $r_T$ is the radius of the semi-wormhole's throat and $r_H$ is the root of $F(r) = 0$ that corresponds to the external event horizon of the relevant black hole, and $d\Omega_{N-2}$ is the usual "surface area" element of an (*N-2*)-dimensional unit sphere. For an *F* corresponding to an asymptotically anti-de Sitter Reissner-Nordstrøm metric, a Wheeler-DeWitt-style treatment of the corresponding thin-shell semi-wormholes suggests that discrete radii exist at which these wormholes are quantum mechanically stable [3, 7, 8]. Envelopment by the brane constituting our universe of these stable, bulk-inhabiting, micro semi-wormholes – also known as "void bubbles" – would induce in our universe the formation of micro wormholes that would be all but indistinguishable from massive particles whose interactions are purely gravitational.

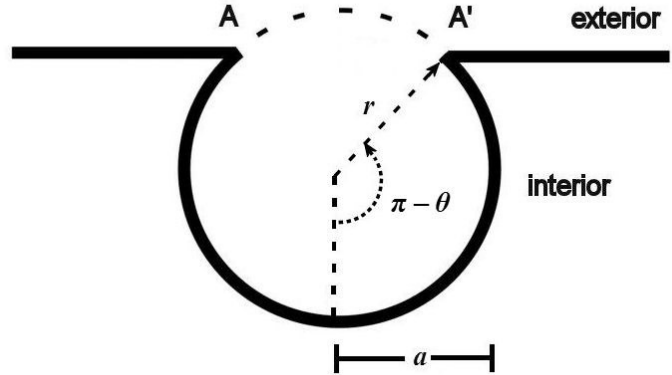

**Figure 1. Partially Enveloped Bulk Wormhole.** Throat of the induced wormhole is at A & A'. Parametric expression for brane embedding depends on *r(s)* and *θ(s),* where *s* (not shown) is radial distance from the point *θ = π*. If *θ > π /2*, the throat ceases to violate the null energy condition, and the induced structure is called a "dimple". [Diagram not intended to depict a realistic embedding that solves the Euler-Lagrange equations.]

Let the dimensionality of the brane be *D*, where *D* < *N*. The coordinates $X^\mu$ of the *N*-dimensional bulk are given by $(X^\mu)=(t, r, \theta_1, \ldots \theta_{N-2})$ and those $\zeta^a$ of the *D*-dimensional brane by $(\zeta^a)=(t, s, \theta_1, \ldots \theta_{D-2})$. The embedding of the spherically symmetric brane within the bulk can be given in terms of the bulk coordinates parametrized by the radial brane coordinate *s*,

$$r = r(s)$$

$$\theta_{D-1} = \theta(s) \tag{2}$$

$$\theta_D = \ldots = \theta_{N-2} = \pi/2$$

Here we deviate slightly from Frolov's treatment [5] (whose notation I have adopted) in order to permit arbitrary spherically symmetric brane configurations, a large class of which his chosen parametrization cannot describe. The metric $\gamma_{ab}$ induced on the brane by the bulk's geometry,

$$\gamma_{ab} = g_{\mu\nu} \frac{\partial X^\mu}{\partial \zeta^a} \frac{\partial X^\nu}{\partial \zeta^b} \quad , \tag{3}$$

has by the definitions of the coordinate systems the line element

$$ds^2 = \gamma_{ab} d\zeta^a d\zeta^b$$
$$= -F dt^2 + \left(F^{-1} r'^2 + r^2 \theta'^2\right) ds^2 + r^2 \sin^2\theta \, d\Omega_{D-2} \tag{4}$$

where $r' \equiv dr/ds$ and $\theta' \equiv d\theta/ds$. On the microscopic length scales of interest (corresponding to quantum bulk wormholes), the gravitational field produced in the bulk by the brane can be neglected. We need not, therefore, impose the usual Darmois-Israel junction conditions [25, 26] in an effort to work out this field. Nor, on this microscopic scale, need we be concerned with the modifications of the on-brane Einstein equations which describe such factors as gravitational brane self-interaction through the bulk [14]. Rather, I shall assume the motion of the brane to be determined by the local gravitational field and topology due to the impinging wormhole, whose exterior spacetime is that of a black hole. Accordingly, we may assume with Frolov [5] that the motion of the brane is determined by the Dirac-Nambu-Goto action [9, 10, 11]

$$S = -T \int d^D \zeta \sqrt{-\det(\gamma_{ab})} \quad , \tag{5}$$

where *T* is the brane tension, which we take to be positive in order to match the positive cosmological constant of our universe. In the static case this action becomes

$$S = -T \Delta t \Omega_{D-2} \int r^{D-2} \sin^{D-2}\theta \sqrt{r'^2 + F r^2 \theta'^2} \, ds \tag{6}$$

where $\Omega_{D-2}$ is the "surface area" of a $(D-2)$-dimensional unit sphere and $\Delta t$ is an arbitrary time interval. This yields the Euler-Lagrange equations

$$Jr'r^{D-2}\sin^{D-3}\theta = 0 \tag{7}$$

$$J\theta'r^{D-2}\sin^{D-3}\theta = 0 \tag{8}$$

where

$$\begin{aligned}J \equiv &-(D-2)r'^3\cos\theta - (D-2)r'Fr^2\theta'^2\cos\theta + (D-2)r'^2 rF\theta'\sin\theta\\ &+(D-2)r^3F^2\theta'^3\sin\theta - Fr^2\theta'r''\sin\theta + \tfrac{1}{2}Fr^4\theta'^3\frac{dF}{dr}\sin\theta\\ &+F^2r^3\theta'^3\sin\theta + \frac{dF}{dr}r'^2r^2\theta'\sin\theta + 2Fr\theta'r'^2\sin\theta + Fr^2\theta''r'\sin\theta\end{aligned} \tag{9}$$

Of course, we are only concerned here with the static case, because we are only interested in the denouement of the encounter. That this "static" case might in fact be quasi-static, given that the absolute stability of the brane is not assured, is of little consequence to the matter under consideration. Specifically, our concern is not with the details of the particular state of the motion of the brane-wormhole system in the arbitrarily distant future. It is merely to determine whether the impinging bulk wormhole necessarily induces the formation of a baby universe through total envelopment by the brane, or whether it comes to rest against the brane – negligibly or substantially (but not totally) enveloped by it. In other words, we will take any period for the onset of instability (presumably determined by the brane tension) to be much greater than the time required for the wormhole-brane encounter to reach its denouement, so that $\Delta t$ in (6) is shorter than the former but much longer than the latter and begins after the denouement is reached. Alternatively, the question of stability can be circumvented by assuming the existence of unspecified stabilizing forces (e.g. those at orbifold fix points[18] or due to the Casimir effect [27] ) in the given braneworld model [5].

As stated above, the wormhole-brane encounter will likely result in several bounces [12] and explosive releases of energy in the form of brane-bound matter fields as well as gravitational waves. These releases, however, together with the positive Keplerian mass of the wormhole guarantee that when the smoke clears, the quantum bulk wormhole (presumably in its ground state) will be adjacent to the macro brane with the latter enveloping the former to an undetermined degree. Our purpose here is not to model the complicated dynamics resulting from speculations on the nature of brane-brane interactions. Rather, we shall only assume that the interaction is not sufficiently violent to destroy the wormhole, that the interaction is dissipative, that the branes cannot pass through each other or coalesce, and that the gravitational attraction of the wormhole to the macro brane will consequently ensure a static (given the assumptions above) end state. These assumptions are consistent with our use of a low-energy theory – general relativity – to describe the encounter.

## III. WORMHOLE ENVELOPMENT

The question of whether total envelopment of the bulk wormhole by the brane is a possible end state of a wormhole-brane encounter is now readily answered. Defining total envelopment by

$$r = a \tag{10}$$

$$\theta = \pi - \frac{s}{a}, \tag{11}$$

where $a$ is the throat radius of the incident bulk wormhole, we see that eq. (7) is immediately satisfied and that eq. (8) requires that $J = 0$. This condition becomes, after inserting eqs. (10) and (11) into (9),

$$\left.\frac{dF}{dr}\right|_{r=a} = -\frac{2(D-1)}{a} F(a). \tag{12}$$

Total envelopment also assumes that the throat of the wormhole induced in the brane has fully constricted to a filament connecting the brane to a spherical pocket universe that surrounds the throat of the bulk wormhole. This filamentary throat may be described by the equation $\theta = 0$ for $r > a$, which clearly satisfies the Euler-Lagrange equations (7) and (8). Because topology change is forbidden within general relativity, I am assuming that the existence of a filamentary throat (i.e. one whose radius is arbitrarily small) signals the presence of a topology change in whichever more permissive and presumably truer theory general relativity serves as a low-energy limit.

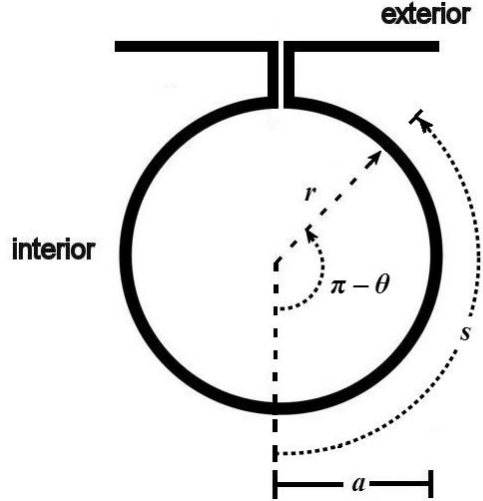

**Figure 2. Totally Enveloped Bulk Wormhole.** Filamentary connection between pocket universe (interior) and the greater brane (exterior) signals the birth of a baby universe.

We require the bulk to be free of matter except at its brane boundaries – the throats of any higher-dimensional bulk semi-wormholes. Accordingly, we form these semi-wormholes by terminating higher-dimensional black hole solutions at a radius $a$ outside of its event horizon. Choosing a spherically symmetric black hole solution consistent within an anti-de Sitter bulk, we have

$$F = 1 - \frac{M_N}{r^{N-3}} + \frac{Q_N^2}{r^{2N-6}} - \Lambda_N r^2 \tag{13}$$

where

$$M_N \equiv \frac{16\pi G_N M}{(N-2)c^2 \Omega_{N-2}} \tag{14}$$

$$Q_N^2 \equiv \frac{k_N Q^2 G_N}{2(N-2)(N-3)} \left( \frac{8\pi}{\Omega_{N-2} c^2} \right)^2 \tag{15}$$

$$\Lambda_N \equiv \frac{2\Lambda}{(N-1)(N-2)} \quad . \tag{16}$$

*M, Q, Λ* the asymptotically observed mass, charge, and cosmological constant. $G_N$ and $k_N$ are respectively the *N*-dimensional gravitational and electrostatic constants, and $\Omega_{N-2}$ is the "surface area" of an (*N-2*)-dimensional unit sphere. Inserting eq. (13) into (12) and specializing to the case of a 3-brane universe (*D* = 4), we have

$$-\Lambda_N a^{2N-4} + \tfrac{3}{4} a^{2N-6} + \frac{N-9}{8} M_N a^{N-3} - \frac{N-6}{4} Q_N^2 = 0 \tag{17}$$

which becomes in the case of an uncharged bulk wormhole,

$$-\Lambda_N a^{N-1} + \tfrac{3}{4} a^{N-3} + \frac{N-9}{8} M_N = 0 \tag{18}.$$

Comparing eq. (18) with the equation for the horizons of the bulk wormhole – namely, *F* = 0, i.e.

$$-\Lambda_N r^{N-1} + r^{N-3} - M_N = 0 \quad , \tag{19}$$

we find that the positive real roots of (18) (for the values of *N* at which they exist) are necessarily smaller than those of (19), the latter corresponding to black hole horizons. In other words, total envelopment is only possible if the throat radius $a$ of the bulk wormhole is smaller than the event horizon of the corresponding black hole. Total envelopment by the brane of an incident bulk wormhole requires the wormhole to be a black hole. Put another way, total envelopment is not a possible static or quasi-static end state of the encounter, because the envelopment necessarily occurs within an event horizon within which the relevant portion of the brane must fully and inexorably contract.

If a bulk wormhole is not a black hole, its envelopment, then, must only be partial. The encounter of the brane with a bulk wormhole must therefore induce the formation within the brane of the structures corresponding to partial envelopment. Cases of partial envelopment may be classified as strong or weak.

Strong envelopment occurs when the throat of the induced structure (the hyperspherical border between the structure's interior and exterior regions, i.e. the $D$-2-sphere containing points $A$ and $A'$ in Figure 1) violates the null energy condition within the brane. In this case the structure is a wormhole whose dimensionality matches that of the brane. This wormhole persists in the sense that it presumably remains in the state of partial envelopment and does not, for the reason adduced, become fully enveloped and pinch off. Were it macroscopic, denizens of the brane would recognize it as a wormhole to a pocket universe. It is microscopic, however, with a radius perhaps on the order of $10^{-22}$ cm and a mass perhaps on the order of $10^4$ TeV [7]. Hence, such an induced wormhole would be perceived instead as an ultra-massive particle, whose interactions are purely gravitational.

Weak envelopment occurs when the induced "throat" actually satisfies the null energy condition. In this case we would describe the persistent structure, not as a wormhole, but as a "dimple". Given that dimples are also due to incident micro wormhole coming to rest against the brane, they also masquerade as particles with of the aforementioned mass and size.

These sorts of gravitationally interacting massive particle-like objects (GIMPs) might serve as a constituent of dark matter (see [13] for a recent review of other dark matter candidates). To ensure a sufficient quantity of these GIMPs, one might suppose the bulk to be awash in tiny semi-wormholes, each of whose throats is coincident with an ($N$-2)-spherical micro brane. Certain of these void bubbles would impinge upon the macro brane that constitutes our universe, become embedded there, and manifest as tiny wormholes or dimples that are perceived by us brane-dwellers as the aforementioned GIMPs.

## IV. DISCUSSION

Although we have shown that persistent induced wormholes and dimples are a theoretical possibility, what evidence can we adduce to justify their inclusion in the list of nonbaryonic cold dark matter candidates? Does not the exotic matter necessarily residing at the throats of wormholes rule them out as realistic solutions to the dark matter problem? First, it is important to remember that the presence of exotic matter is not a violation of the laws of physics. Exotic matter – violations of the null energy condition – are well known to occur in certain quantum systems, the most familiar example being the system consisting of two Casimir plates. Hence, an exotic matter component within the quantum wormholes induced in our brane may not be ruled out, as it is completely consistent with ordinary physics.   Recall also that the metric in the empty region beyond the immediate vicinity of a wormhole's throat is, by a practical extension of Birkhoff's theorem to the Einstein equations on the brane [14], that of a black hole in the region exterior to its outer event horizon.  In other words, the Keplerian effects of an exotic-matter-containing quantum wormhole would in general be identical to that of an ordinary massive particle.  [Although it is in principle possible for a wormhole to possess a negative Keplerian mass, this bizarre possibility is seldom considered in the literature, and I have not considered it here.]  Lastly, recall that the throat of a bulk wormhole will automatically violate the null energy condition if it is taken, as I have taken it here, to be a negative-tension brane [3] -- a spatially closed version of an entity commonly featured in brane world scenarios since their inception over a decade ago [2].  Dimples, in contrast to wormholes, are unproblematic in that they do not violate the null energy condition.

The GIMPs described here may be likened to another conjectured cold dark matter candidate that also interacts purely through gravitation -- primordial black holes.  Unlike the wormholes under consideration, however, these black holes are necessarily macroscopic and classical objects.  Were they instead microscopic and quantum, they would immediately evaporate away.  By contrast, the quantum wormholes and dimples described above, lacking event horizons, do not emit Hawking radiation.  They can persist indefinitely.  In short, primordial quantum black holes cannot exist, but primordial quantum wormholes can.   Here I have supposed further that the latter might have formed in our brane as a consequence of collisions with semi-wormholes (void bubbles) in the bulk.

It is important to emphasize the motivation for considering the interaction between our brane and bulk-inhabiting wormholes, as opposed some type of unknown bulk-inhabiting particles.  It is, as previously stated, the aforementioned brane world proscription of off-brane matter.  Although one can ignore this proscription and explicitly consider bulk-dwelling matter – a dilaton field (see, for example, [15]) or, more egregiously, a matter fluid in the bulk [16, 12, 17] -- one must then explain why certain types of matter are confined to the brane while others are not.  Wormholes with branes at their throats and brane-bounded void bubbles (semi-wormholes) are, by contrast, very simple, fully localized, bulk-inhabiting objects consistent with the brane world ethos.  Moreover, they are the only localized objects required by (higher-dimensional) general relativity to violate the null energy condition – i.e. to have negative tension.  Placing these branes, then, at the throats of wormholes is analogous to the usual practice of situating negative-tension branes at orbifold fixed points [see, for example, 18].

The choice of bulk wormholes – or more precisely, brane-bounded void bubbles -- is further motivated by the original Randall Sundrum conception of branes as boundaries of the bulk.  Hence, the treatment here differs markedly from that of Gen, Ishibashi, and Tanaka [19] who considered the collision between our 3-brane universe and a "vacuum bubble".   The latter arises through spontaneous nucleation when the bulk is assumed to be in a false vacuum state.  Unlike void bubbles, vacuum bubbles enclose a lower energy vacuum (the "true" one), are not necessarily surrounded by a brane, and are explosively expanding. In other words, vacuum bubbles, unlike void bubbles, are not stable boundaries between the bulk and literal nothingness.

The increase in the dark matter content of our 3-brane universe due to the ongoing accretion of bulk wormholes will have cosmological consequences. These can be gleaned from recent models of brane world cosmology that permit the transfer of mass-energy from the bulk or explicitly define dark matter as matter transferred from the bulk [20, 21, 16, 17]. That these models are consistent with the observed properties of the universe supports the notion that bulk-to-brane matter transfers – including those resulting from the capture of bulk wormholes – are within the realm of possibility. A purpose of this note has been to suggest a means, through which this transfer could have occurred, that is consistent with the original brane world hypothesis in that it does not require us to assume the existence of matter within the bulk.

## V. OBSERVATIONAL CONSEQUENCES

The observational consequences of these conjectured quantum wormholes are difficult to quantify. Because induced quantum wormholes are not thermal relics of the big bang and are not even particles, the standard calculation of the relic density [13] does not apply. A stubborn attempt to perform this calculation would face steep challenges. It is unclear, for example, whether induced wormholes and dimples are effectively annihilated when they collide. Nor, in the absence of a fully formulated brane world cosmology featuring a void-bubble-containing bulk, can we estimate the rate of induced wormhole and dimple production. Induced wormholes and dimples, then, most closely resemble light wimpzillas with a likely annihilation cross section of zero, an unknown production cross section, and a nonexistent weak-interaction coupling to other matter.

Such a wormhole candidate leads one to expect less dark-matter clumping than in the usual WIMP-base models. The reasons are clear. The mass of a quantum wormhole is of the (higher dimensional) Planck scale [7], orders of magnitude larger than the lightest supersymmetric WIMP [13, 22]. The number density of wormholes is correspondingly smaller, which would result in far fewer collisions between wormhole GIMPs than between supersymmetric WIMPs. Moreover, WIMPs are subject to the weak interaction, while GIMPs may only interact gravitationally. This lessened mutual interaction would also presumably result in reduced scattering and therefore less clumping. This in turn would result in galactic dark-matter halos of greater spatial extent with apparent gravitational lensing farther beyond luminous galactic boundaries. Hence, unexpectedly ubiquitous gravitational lensing in the absence of nearby luminous matter together with a failure of the LHC to find neutralinos or other supersymmetric particles, could possibly favor induced quantum wormholes or other GIMPs as candidates for nonbaryonic dark matter. Such dark matter candidates, members of the category of non-WIMP explanations, have recently become more interesting in light of the negative experimental findings of the Cryogenic Dark Matter Search (CDMS) collaboration [23]. Although the failure of CDMS to detect any WIMP interactions might intensify the search for non-WIMP particle candidates (see for example [24]), it might be time as well to consider seriously dark matter candidates, such as induced quantum wormholes, that are not particles at all.

## VI. CONCLUSION

To summarize, horizon-free bulk wormholes that encounter the brane presumed to constitute our universe cannot become totally enveloped by it. The encounter will not, therefore, result in the birth of a baby universe. Instead, envelopment by the brane will be partial and will thereby manifest as induced dimples or wormholes that persist. Because of their microscopic scale and purely gravitational interactions, these objects would be perceived as particles of dark matter. Computing the corresponding density of dark matter, for comparison with the value inferred from current observations, can only proceed in the context of a suitably formulated brane-world cosmology.

.